\documentclass[a4paper,11pt]{article}
\pdfoutput=1 

\usepackage{jinstpub} 

\usepackage{graphicx}\usepackage{dcolumn}\usepackage[mathlines]{lineno}\usepackage{xspace}
\usepackage[nolist]{acronym}
\usepackage[dvipsnames]{xcolor}
\usepackage{caption}
\usepackage{subcaption}
\usepackage[inline]{enumitem}
\usepackage{comment}
\usepackage{bm}
\usepackage[normalem]{ulem}

\graphicspath{{images/}}

\title{Muon Track Reconstruction in a Segmented Bolometric Array Using Multi-Objective Optimization}

\author[1]{J. Yocum,\note{Corresponding author.}}
\author{D. Mayer,}
\author{J. L. Ouellet,}
\author{and L. Winslow}

\affiliation{Laboratory for Nuclear Science, Massachusetts Institute of Technology, Cambridge, MA 02139, USA}

\emailAdd{juliany@mit.edu}
\emailAdd{lwinslow@mit.edu}

\abstract{Recent advances in segmented solid-state detector arrays for rare-event searches have allowed the technology to approach the ton-scale in detector mass and the scale of meters in size.  Often focused around searches for neutrinoless double-beta decay or direct dark matter detection, such experiments also have the capability to search for exotic particles that leave track-like signatures across their volume.  However, the segmented nature of such detector arrays often sets the spatial resolution and makes the problem of reconstructing track-like paths non-trivial. In this paper, we present an algorithm that improves reconstruction of track-like events in segmented detectors using multi-objective optimization -- a computational technique that optimizes more than one cost function at a time without specifying a quantitative weighting between them. Such a technique allows the reconstruction of tracks through a detector and the determination of path-lengths through individual elements. When combined with the reconstructed energy depositions in each element this allows for a calculation of the stopping power of track-like particles and opens the door to searches for particles with abnormal stopping power like monopoles or lightly-ionizing particles (LIPs). Results are presented which evaluate the precision of the reconstruction tools as they currently stand against Monte Carlo generated data. The algorithm is presented in the context of the CUORE experiment, but has applications to other segmented calorimeter detectors. 
}


\begin{document}
\maketitle
\flushbottom
\acresetall

\begin{acronym}
\acro{SM}{Standard Model}
\acro{BSM}{Beyond Standard Model}
\acro{NDBD}[0\ensuremath{\nu\beta\beta} decay]{neutrinoless double beta decay}

\acro{CUORE}{Cryogenic Underground Observatory for Rare Events}
\acro{CUPID}{CUORE Upgrade with Particle Identification}

\acro{NTD}{neutron transmutation doped}
\newacroindefinite{NTD}{an}{a}

\acro{LNGS}{Laboratori Nazionali del Gran Sasso}

\acro{MOO}{Multi-Objective Optimization}
\acro{PDF}{probability density function}
\acro{MC}{Monte-Carlo}
\acro{LSQ}{Least Squares}

\acro{LIP}{lightly-ionizing particle}
\acro{MIDM}{multiply-interacting dark matter}

\end{acronym}

\newcommand{\ndbd}{\ac{NDBD}\xspace}
\newcommand{\kgyr}{\ensuremath{\mathrm{kg\cdot yr}}\xspace}
\newcommand{\ckky}{\ensuremath{\mathrm{cnts/(keV\cdot kg\cdot yr)}}\xspace}
 
\section{Introduction}

Track-like event topologies are no stranger to particle detectors. Muon tracks within cloud chambers were one of the first direct observations of particle interactions.  In recent years, segmented arrays of large solid-state detectors searching for \ac{NDBD}, dark matter, or other rare-events have reached suitable sizes so as to enable particle tracking across their volume.  Despite their smaller exposures in terms of area-steradian coverage compared to traditional large-volume liquid or plastic scintillators or gaseous trackers, such contemporary solid-state detectors benefit from high energy resolution and low $dE/dx$ thresholds. Combined with their typical placement in deep underground labs, this can result in competitive sensitivity to exotic track-like phenomena including multiply-interacting dark matter \cite{PhysRevD.102.123026} and \acp{LIP} \cite{PhysRevLett.127.081802, PhysRevLett.120.211804}. Conversely, as the background requirements of these rare-event searches become increasingly stringent, it is valuable to have a precise \textit{in situ} characterization of cosmogenic backgrounds such as muon-induced spallation products \cite{cosmogenic1,cosmogenic2,cosmogenic3}.  Either in tandem with dedicated muon-veto detector systems or as a stand-alone reconstruction, precise tracking information of muons crossing the detector volume can enable data-driven studies of these backgrounds. 

However, as these segmented detector arrays were not designed for particle tracking, many of their features make precise track reconstruction difficult. Chief among these are poor spatial and temporal resolution. In rare event searches, many radioactive backgrounds are distributed on the surfaces of detector modules, while sensitivity scales with the module volume. As a result, rare-event searches tend to opt for arrays of large detector modules with large volume-to-surface ratios. The detectors have poor spatial resolution (of the order of several cm) and poor timing resolution (of the order of 10's of $\mu$s to 100's of ms). For such detector arrays, it is imperative to exploit as much geometric information as possible when reconstructing detector-wide events in order to maximize the physics reach. Potential searches for new physics, along with in-depth cosmogenic background characterization, motivate the development of track reconstruction algorithms tailor-made for these segmented solid-state detectors. 

In this paper, we present an algorithm using multi-objective optimization for the purpose of reconstructing track-like events within a large segmented solid-state detector array with low spatial and temporal resolution. Multi- and many-objective optimization techniques allow for one to find solutions to problems which must simultaneously extremize over multiple objective functions without needing to specify weights \textit{a priori}. The algorithm presented here is implemented using the \texttt{pymoo} library, with objective functions and problem constraints chosen to reflect the physics of track-like energy depositions within the geometry of the existing \acs{CUORE} detector. Though the algorithm presented here can be applied to any particle (Standard Model or beyond) that leaves a linear track, we will mainly focus on muons throughout this paper as they are the most common expected track-like interaction in \acs{CUORE}.

\section{The CUORE Detector}   

\begin{figure}[!ht]
\centering
\includegraphics[trim=800 0 800 0, clip, height=.35\textwidth]{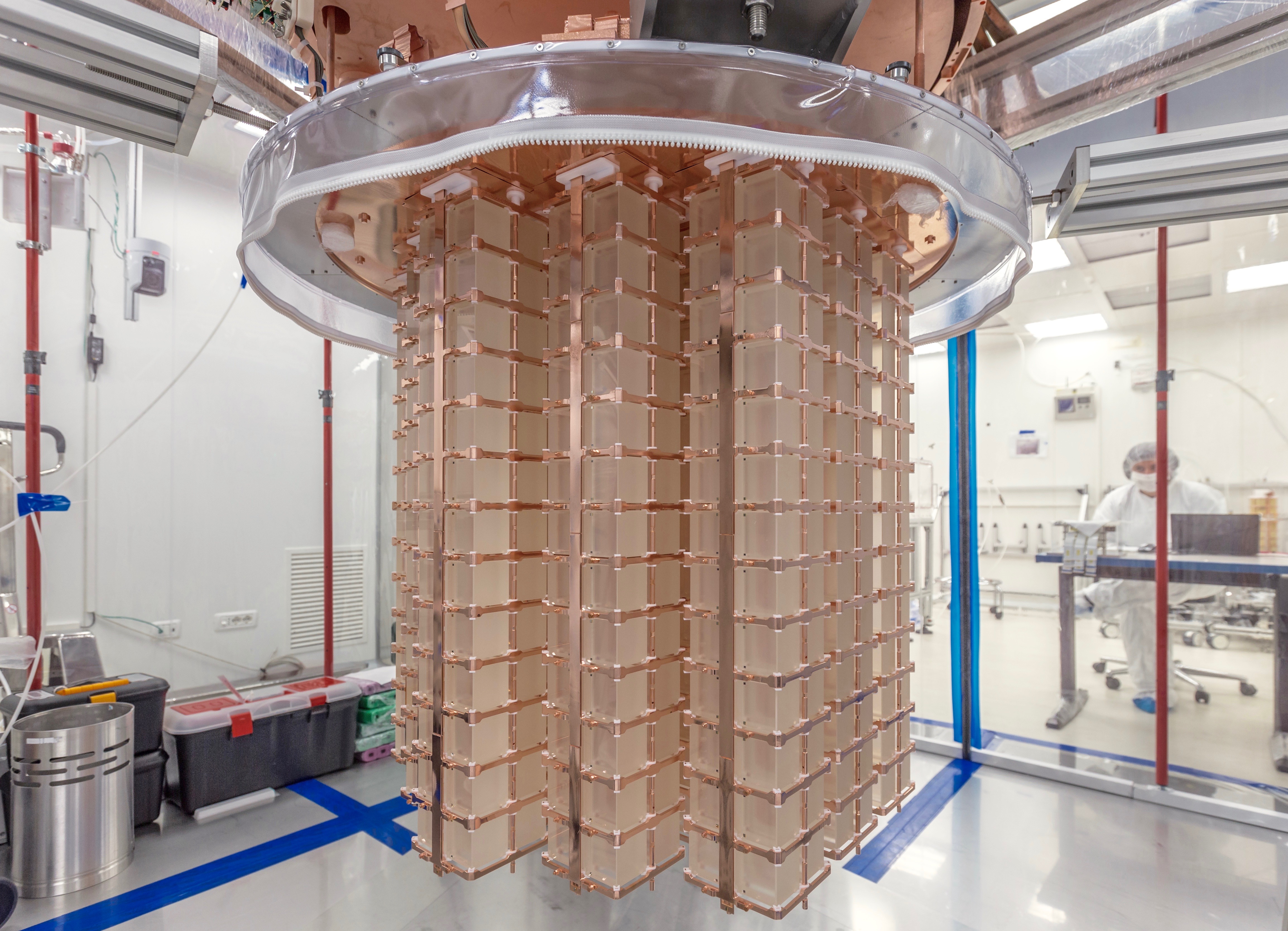}
\hspace{20mm}
\includegraphics[height=.35\textwidth]{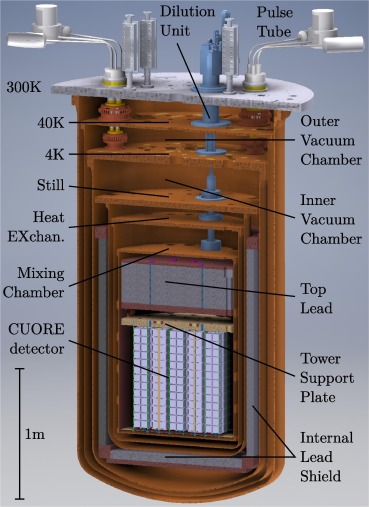}
\caption{\emph{Left:} The CUORE detector suspended from the CUORE cryostat during detector installation. The crystal absorbers are suspended in copper frames in a 19-tower, 13-floor array. \emph{Right:} A schematic of the CUORE detector inside the CUORE cryostat. The detector is surrounded by nested layers of copper thermal shielding and lead $\gamma$ shielding. An additional 24\,cm thick lead shield (not pictured) surrounds the bottom and sides of the cryostat. Figures from \cite{CUORE-Cryostat,CUOREWebpage}.}
\label{fig:cuore_pictures}
\end{figure}

The \ac{CUORE} is a ton-scale cryogenic detector searching for \ac{NDBD} in $^{130}$Te \cite{CUORE-300kgyr,CUORE-2NuHalfLifePaper,CUORE-OneTonYear}. Situated within the \ac{LNGS}, the detector is composed of 988 high-purity 5\,cm$\times$5\,cm$\times$5\,cm TeO$_2$ crystals, each instrumented as a macroscopic cryogenic calorimeter. The crystals are arranged into 19 towers of 13 floors each, providing a segmented geometry which allows for the reconstruction of energy depositions occurring within multiple crystals and across the detector (see Fig.~\ref{fig:cuore_pictures}). An energy deposition in any crystal leads to a sudden rise in temperature (on the order of $100\,\mu\mathrm{K/MeV}$) \cite{CUORE-OneTonYear}, which can be triggered and used to extract the energy deposited. Each crystal is operated independently, but can be combined offline to look for simultaneous energy depositions across multiple calorimeters. 

The detector is housed within the \ac{CUORE} cryostat \cite{CUORE-Cryostat} and held at a temperature of approximately 12 mK. Ancient roman lead shielding within the cryostat efficiently blocks external gamma rays from reaching active detector components, while stringent materials screening and selection during CUORE's construction serve to reduce intrinsic radiogenic backgrounds. CUORE has collected more than 1000\,\kgyr in exposure, to-date the largest exposure of a cryogenic solid-state based detector.

Each crystal in CUORE is thermally isolated, save for a weak thermal link to a cold bath at $\approx12$\,mK, and \iac{NTD} thermistor \cite{Haller1984}, which acts as a sensitive thermometer.  Energy depositions within the crystal will yield a heat deposition, which is read out as a voltage pulse on the \ac{NTD} with risetimes of $\sim50$\,ms and falltimes of $\sim1$\,s. The voltage from each calorimeter is digitized at 1\,kS/s and stored to disk as a continuous waveform. Pulses are triggered offline using an optimum-filter based trigger \cite{Arnaboldi_2018,Domizio_2018}, and are analyzed to extract the pulse amplitude, stabilize against thermal gain drifts, and calibrate the amplitude into an estimate of the deposited energy \cite{CUORE0-AnalysisTechniques}. During physics data taking, individual crystals have typical trigger rates of a few mHz. 

Finally, pulses are are time-correlated and classified according to their multiplicity, that is the number of calorimeters which are found to have triggered in coincidence with each other. The typical timing resolution of two simultaneous energy depositions is on the order of 5-10\,ms. Track-like muons within CUORE may be expected to give rise to multiplicities between 1 to $\sim$20, depending on their geometric intersection across the detector. The long thermal timescales involved in the calorimetric technique are unable to resolve time-of-flight measurements for relativistic particles traversing the detector.  Practically, all prompt events induced by a given muon within CUORE will appear simultaneous, up to the timing resolution \cite{PhysRevC.93.045503}.  Nonetheless, the low background rates within the detector lead to >90\% of muon tracks being free from pile-up with accidental coincidences within the detector.

As CUORE is optimized for its energy resolution about the $Q_{\beta\beta}$-value for $^{130}$Te of 2.528\,MeV \cite{PhysRevLett.102.212502,PhysRevC.80.025501,RAHAMAN2011412}, single-crystal depositions of  $\gtrsim20$\,MeV are found to saturate the digitizer dynamic range. As muons can deposit energies well above this range, this presents a challenge for muon reconstruction algorithms which rely on the reconstructed energy. Future efforts may seek to recover such saturated pulses, through the use of the leading and trailing edges of the pulse in conjunction with the known thermal response. Conversely, energy depositions close to a calorimeter's trigger threshold may fail to generate an event and could result in energy missing from a track. While this may become important in searches for exotic particles, it is rarely a problem for muons, which tend to leave energy depositions well over the trigger thresholds, even as minimum ionizing particles. 

The large overburden provided by \ac{LNGS} attenuates the vast majority of the cosmic muon flux. The surving flux has been measured to be $\Gamma\approx(3.2\pm0.2)\times10^{-8}\,\mathrm{\mu\cdot s^{-1}\cdot cm^{-2}}$ \cite{MACRO:1993jdl,PhysRevD.58.092005}, which implies a muon interaction rate in CUORE of approximately 1 per hour \cite{BELLINI2010169}. The distribution of muons is peaked about the vertical, with almost no flux with zenith angle $\cos\theta<0.4$ and an average of $\langle\cos\theta\rangle\approx0.8$. The average muon energy is $\langle E_\mu\rangle\approx270$\,GeV \cite{MACRO:1993jdl,PhysRevD.58.092005}.  Muons which have reached \ac{CUORE} -- and have not interacted in or near the detector so as to induce an electromagnetic shower -- will deposit energy in a track-like manner as they traverse the crystals of the detector.  The energy deposited in an individual crystal will be related to both the track-length travelled, along with fluctuations intrinsic to the processes of energy loss within media.  Muons reaching \ac{LNGS} are typically close to minimally-ionizing particles , with most probable energy losses on the order of a few $\mathrm{MeV\cdot g^{-1}\cdot cm^{-2}}$.

\section{Multi-Objective Optimization}
\label{sec:moo}

The segmented nature of the CUORE detector makes track reconstruction difficult. The detector is relatively small and with coarse spatial segmentation compared to a typical muon tracker. A naive \ac{LSQ} algorithm, which minimizes the sum total of squared distances between a reconstructed track and the centers of crystals with energy depositions tends to be biased towards pulling the reconstructed track towards the crystal centers rather than toward the true track. \ac{LSQ} is susceptible to outliers, for example an accidental (uncorrelated) coincidence which may be far from the track and pull the reconstructed track. It also does not utilize additional information such as the positions of crystals which did \emph{not} detect any energy, or the amount of energy deposited in each crystal. A more robust track reconstruction algorithm would take advantage of all the available information and treat the detector as a whole rather than focusing only on the channels where energy was detected. 

The algorithm that we propose here attempts to find a track that optimizes the channels where energy is detected, the ``lit channels,'' the channels where no energy is detected, the ``unlit channels,'' as well as the best estimate of the energy that was deposited in each lit channel. Defining a generic optimization of these three features in any given event is non-trivial due to the geometry of the CUORE detector. For example, it is not obvious how to quantitatively weight fitting the lit channels in any particular event against fitting the unlit channels. This weighting may depend on the direction of the track and the path the track takes through the detector. In this work we do not attempt to model it.

Multi-objective optimization describes a class of optimization techniques which aim to optimize several objective functions simultaneously. This is often useful when the relative importance between the objectives are uncertain or poorly defined. By optimizing simultaneously, a multi-objective optimization process searches the space of possible trade-offs between the objectives. Often, the outcome of an optimization of $n$ objectives is an $(n-1)$-dimensional ``Pareto front'', defining the subspace of optima for all possible weightings of the defined objectives.

In the algorithm presented here, we found that minimizing over three particular cost functions was able to reproduce simulated tracks with good reliability. The motivation for each of the functions came from a corresponding property of the track to be optimized for. The first two cost functions are meant to only consider geometrical information for characterizing a track. The first two objective functions assign costs on the basis of whether a candidate trajectory has intersected as many “lit” crystals as possible (in the case of the first objective), or has minimized the number of “unlit” crystals as much as possible (in the case of the second objective). The first two objectives are designed to be computationally efficient approximations to crystal intersections by approximating a crystal intersection with a logistic function, defined separately for ``lit'' and ``unlit'' crystals. The resulting objectives take the form
\begin{equation}
\label{eqn:logistic_cost}
f_{1,2}(x) = \sum_{i\in\Omega_{1,2}}\frac{1}{1+e^{\lambda_{1,2}(x_i-\alpha_{1,2})}}.
\end{equation}
where $x_{i}$ is the minimum distance between the $i^{\rm th}$ crystal center and the reconstructed track, $\alpha_{1,2}$ encodes a geometric ``crystal size'' (described below), and $\lambda_{1,2}$ defines the ``sharpness'' of the crystal edge.

The first objective encourages track collisions with all "lit" crystals. The parameters are given by $\lambda_1=-.2$\,mm$^{-1}$ and $\alpha_1=25\sqrt{3}$\,mm, where $25\sqrt{3}$\,mm is an upper bound on the distance within which a collision could occur (the distance between a crystal center and corner), such that the logistic function returns a penalty close to 1 if the track lies outside of the crystal, and close to 0 penalty if inside. The sum $i$ is taken over all ``lit'' crystals, $\Omega_1$.

Conversely, the second objective function encourages avoidance of track collisions with all ``unlit'' crystals by taking an average of crystal costs and setting $\lambda_2=-.2$\,mm$^{-1}$ and $\alpha_2=25$\,mm, where $25$\,mm is a lower bound on the distance from which a track misses a crystal. The sum $i$ is taken over all ``unlit'' crystals, $\Omega_2$. For both objectives, $\lambda_{1,2}$ was chosen phenomenologically to give reasonable algorithm performance, but may be further tuned in the future to improve performance. 

Lastly, the third objective function encourages $dE/dx$ values with high probability given a particle of interest. The geometric path length through each intersected is calculated by performing line-plane intersections. The probability densities are then computed from a pdf given by GEANT4, from which a log-likelihood based cost is returned:
\begin{equation}
\label{eqn:dEdxPdf}
f_3(x) = -\sum_i \log(p(\beta_i))
\end{equation}
where $\beta_i$ gives the calculated $dE/dx$ of the candidate trajectory through the i$^\mathrm{th}$ crystal based in its reconstructed energy $E_i$ and path length $l_i$,
\begin{equation}
\label{eqn:dEdx}
\beta_i = E_i/L_i,
\end{equation}
and $p$ gives the \ac{PDF} of $\beta$ for a given particle (see Fig.~\ref{fig:dEdx_by_particle}). The sum $i$ currently runs over crystals that are intersected by the proposed track with non-zero energy depositions. In future versions of the algorithm, False-Hits and False-Misses could be incorporated into this sum.

Once these objectives are defined, a two tiered optimization process begins, utilizing two algorithms from Multi-objective Optimization python library \texttt{pyMOO} \cite{pymoo}: NSGA-II \cite{996017} and NSGA-III \cite{6600851,6595567,10.1007/978-3-030-12598-1_19}. NSGA-II (for two objectives) and NSGA-III (for three or more) are genetic algorithms, beginning with some initial population of size $n$ and evolving population members where survival is determined from the objective costs, forming the Pareto front. The algorithm terminates after the completion of a number of evolutionary steps. The first tier begins with NSGA-II, optimizing trajectories over $f_1$ and $f_2$. The initial population is defined as a random sample over the sample space, where trajectories over the sample space are constrained to intersect with at least two crystals which are closest to the \ac{LSQ} fit line. The outcome is a Pareto Front representing the best trajectory candidates by strictly geometrical considerations, which is a sample of those trajectories which intersect with the most "lit" crystals, the fewest "unlit" crystals, and the range of trade-offs in between.

Members of the Pareto front produced from NSGA-II are then defined as the initial population for which NSGA-III will optimize, but now with consideration of $f_3$ in addition to $f_1$ and $f_2$. Once termination is reached, a selection from the Pareto Front is made according to the following scheme:
\begin{enumerate}
\item Compute which crystals were intersected for each candidate track using the line-cube collision.
\item Keep the track candidates which intersect the most ``lit'' crystals;
\item Among these, keep only the track candidates that intersect the fewest number of ``unlit'' crystals;
\item Select from remaining according to be lowest $f_3$ cost (picking randomly for tie-breaking).
\end{enumerate}

Multi-objective optimization of these cost functions together provide a number of advantages over an \ac{LSQ} algorithm. First, the approach is more robust to outliers, such as accidental coincident events. For instance, a randomly coincident evident located far from a trajectory could pull the estimated trajectory given by an LSQ to be closer to that of the coincident event, whereas the cost of the given objective functions will have no effect on the MO optimization, uniformly raising the costs of all track candidates by a constant. Moreover, much more information is utilized in the fitting of a track, such as the geometries of crystals and the overall array, as well as information about energy depositions which would be neglected in a naive \ac{LSQ} fitting. 

Lastly, by optimizing over $f_3$, tracks whose energy deposition profile has high likelihood given its corresponding $dE/dx$ \ac{PDF} are searched for. The likelihood from the best tracks can then be used to perform hypothesis testing on the type of particle detected. This is discussed further in Sec.~\ref{sec:discussion}. 

In experimental data we expect that secondary scattered particles will occasionally deposit energy in crystals that are near to, but not intersected by the primary particle trajectory. These secondary crystals are not accounted for in the current fit, but could be accounted for in future versions. It is worth pointing out that the MOO algorithm is robust to outlier crystals, so the impact on track reconstruction is expected to be small. Large showering events, such as a muon that produces a hadronic shower that are not well described as a track are rare, and are beyond the scope of this work.

\section{Simulation of Muon Tracks}
\label{sec:simulations}

In this algorithm, we use two \ac{MC} simulations: a toy \ac{MC} to quickly calculate trajectories and geometric intersections, and a GEANT4-based \ac{MC} \cite{Geant4} to understand the energy deposition within a single crystal.

\begin{figure}
    \centering
    \begin{subfigure}[b]{0.45\textwidth}
        \centering
        \includegraphics[height=0.8\textwidth]{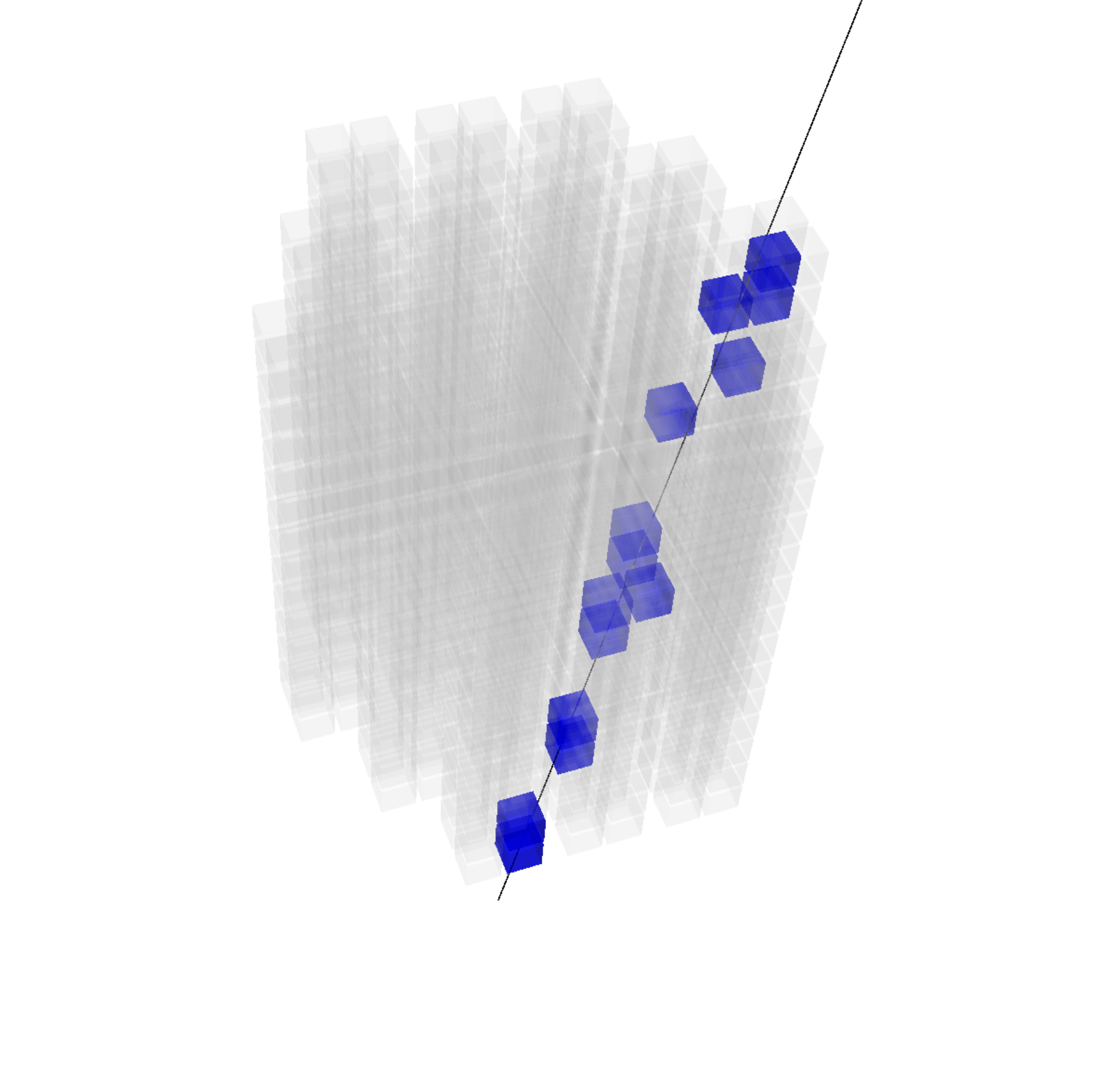}
        \caption{}
        \label{fig:toymc_cuore}
    \end{subfigure}
    \hspace{.5cm}
    \begin{subfigure}[b]{0.45\textwidth}
        \centering
        \includegraphics[height=0.8\textwidth]{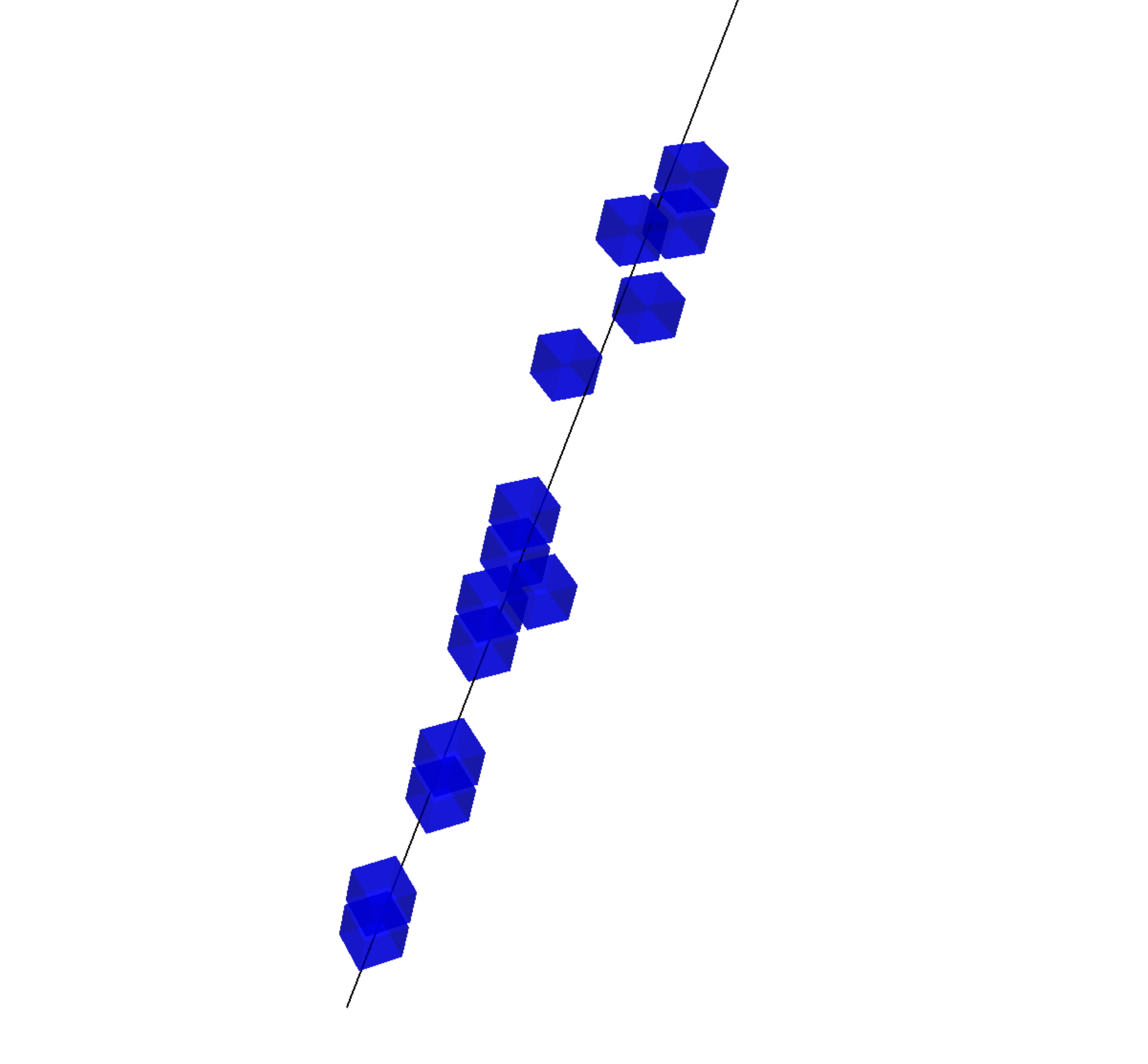}
        \caption{}
        \label{fig:toymc_cuore_hitsonly}
    \end{subfigure}
    \caption{Simulated particle trajectories from the toy \ac{MC}. Left: Trajectory through the entire CUORE array. Intersected crystals are highlighted in blue. Crystals are rendered transparent for clarity. Right: Same trajectory but isolating only the intersected crystals.}
    \label{fig:toymc}
\end{figure}

Since each step of the optimization requires testing the crystal intersections of a proposed track, we designed a toy \ac{MC} to simulate tracks through the CUORE detector. We construct the CUORE detector geometry, including the coordinates of each crystal and its dimensions, excluding support structures and shields. We simulate a track by defining an origin point within the detector and a direction. Each track may contain intersected crystals (e.g. shown in Fig.~\ref{fig:toymc}). Given the chosen trajectory, we then calculate the geometric path length through each intersected crystal. During the optimization, the simulation stops here and the path lengths are used to calculate the third objective function.

To validate our optimizer, we use the toy \ac{MC} simulation to build a population of tracks to fit. In this case the toy \ac{MC} creates a large sample of tracks and takes the additional step of converting path lengths into energy depositions. The energy deposition in a crystal is chosen by randomly sampling the $dE/dx$ \ac{PDF} of that particle and multiplying by the path length. The $dE/dx$ \ac{PDF} is taken from the Geant4 simulation described below. The toy \ac{MC} simulations does not generate secondary showering particles, which are discussed further below.

\begin{figure}
  \centering
  \includegraphics[width=.45\textwidth]{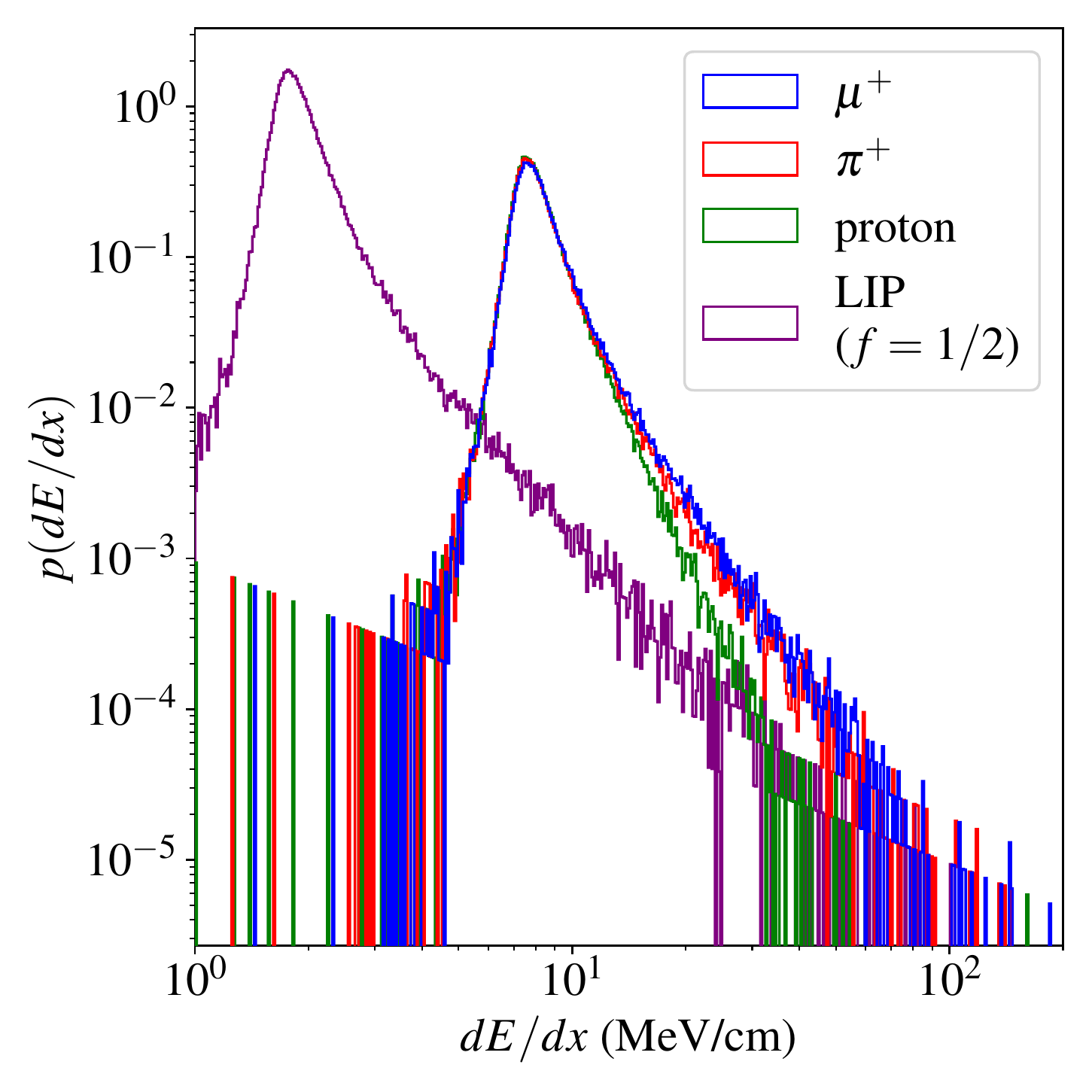}
  \caption{Distribution of simulated $dE/dx$ values in a $5\,\mathrm{cm}\times5\,\mathrm{cm}\times5\,\mathrm{cm}$ TeO$_2$ crystal for 300\,GeV muons, protons, pions, and a hypothetical muon-like \acs{LIP} with charge $e/2$. (\acs{LIP} simulation used modifications to Geant4 from the \texttt{FCP\_Simulation} package \cite{BANIK2020164114,FCP_Simulation}.) Interestingly, an individual crystal of this size is not able to distinguish between muons, pions and protons via $dE/dx$.}
  \label{fig:dEdx_by_particle}
\end{figure}

A GEANT4 simulation was used for calculating $dE/dx$ curves by sampling trajectories of particles of interest through a single 5\,cm $\times$5\,cm$\times$5\,cm cube of TeO$_2$. A particle is simulated at a random point on the surface with a random direction into the crystal. For a given trajectory, we calculate the path length it took through the crystal, as well as the total energy deposited in the crystal over the path, allowing us to approximate $dE/dx$ as in Eqn.~\ref{eqn:dEdx}. Performing this calculation over all sample trajectories, we are left with a $dE/dx$ distribution shown in Fig.~\ref{fig:dEdx_by_particle}.

\section{Algorithm Performance on Muon Tracks}

We test our algorithm's performance on a sample of 1382 muons generated using the MC simulations described in the previous section. We compare our algorithms performance to the performance of a naive \ac{LSQ} algorithm as a benchmark.

We assess the algorithm's ability to accurately reconstruct the original trajectory of a particle to which it was blind. The algorithm takes only the simulated energy depositions, as it would in the case of real data. As mentioned, one motivation for the multi-objective approach was to include distinct objectives for considering crystal collisions in the trajectory. The consideration of collisions creates two categories of error crystals: 1) False-Hits: crystals which were intersected by the reconstructed trajectory but which did not detect any energy depositions. 2) False-Misses: crystals which detected energy depositions but were not intersected by the reconstructed trajectory. 

\begin{figure}
    \centering
    \begin{subfigure}[b]{0.45\textwidth}
        \centering
        \includegraphics[width=\textwidth]{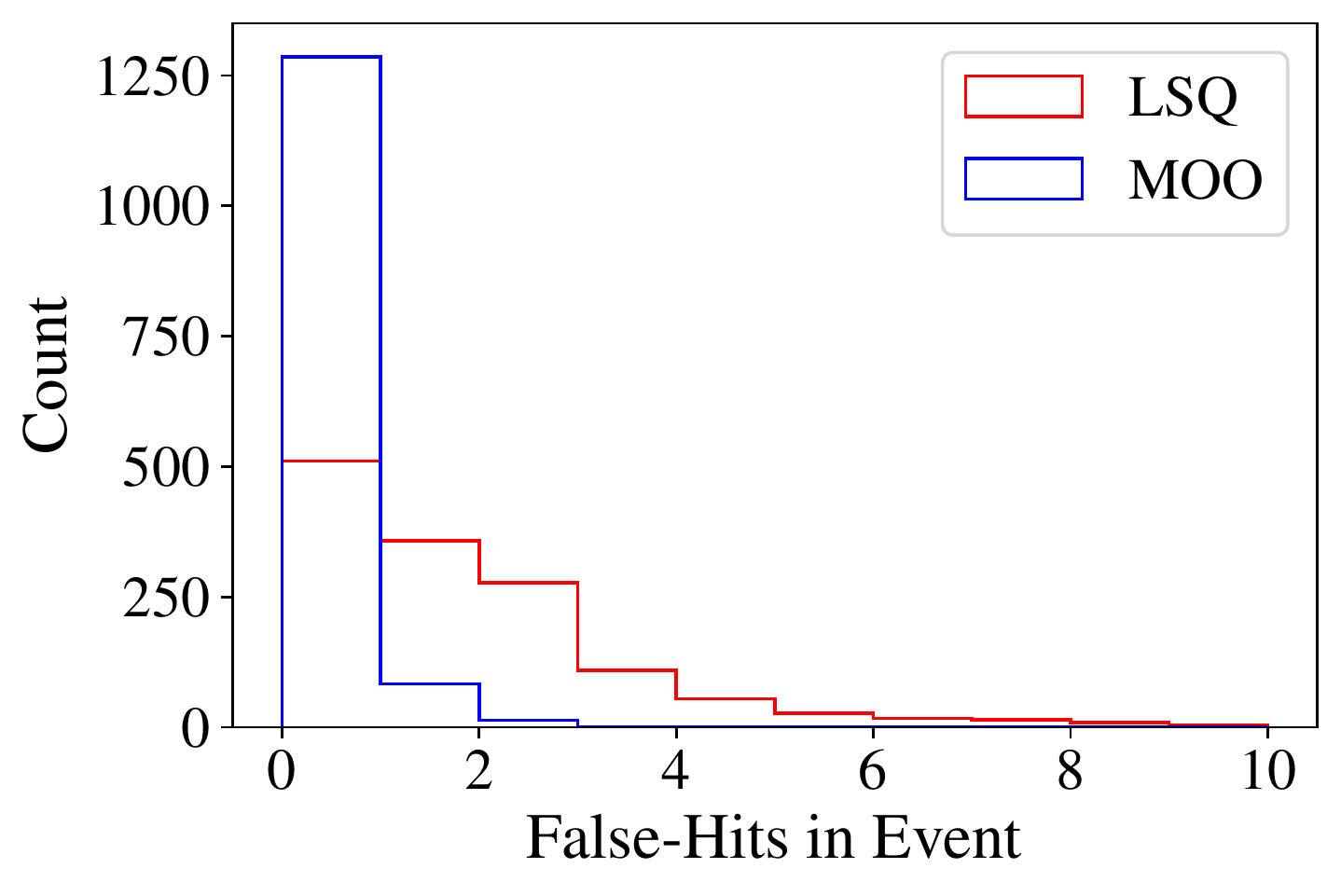}
        \caption{}
        \label{fig:falsehits}
    \end{subfigure}
    \hspace{1cm}
    \begin{subfigure}[b]{0.45\textwidth}
        \centering
        \includegraphics[width=\textwidth]{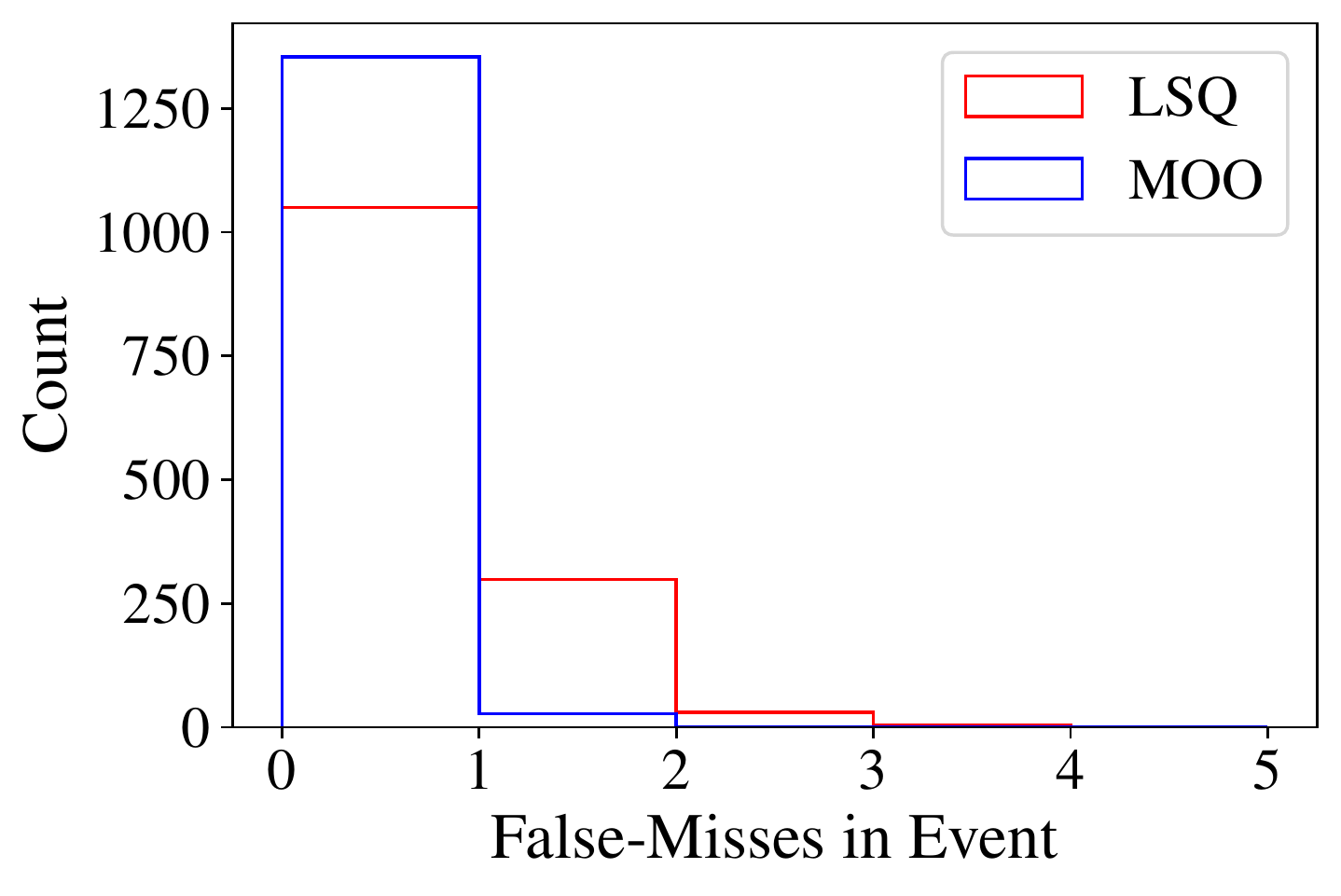}
        \caption{}
        \label{fig:falsemisses}
    \end{subfigure}
    \caption{Distribution of error crystals for \ac{MOO} compared to least squares. Left: Distribution of number of false-hit crystals. Right: Distribution of false-missed crystals. The MOO algorithm out performs the naive LSQ algorithm in terms of both false-hit and false-missed crystals.}
    \label{fig:error_crystals}
\end{figure}

In Fig.~\ref{fig:error_crystals}, we show a comparison between performance of a naive \ac{LSQ} against the multi-objective algorithm over False-Hits and False-Misses. On this sample our algorithm was able to increase the number of reconstructions with no False-Hits from 37\% to 93\%. It was similarly able to increase the number of reconstructions with no False-Misses from 76\% to 98\%. These statistics are particularly important for delayed coincidence analyses to look for cosmogenic crystal activation, which rely on an accurate accounting of which crystals have been struck. 

\begin{figure}
  \centering
  \begin{subfigure}[b]{0.45\textwidth}
    \centering
    \includegraphics[height=0.8\textwidth]{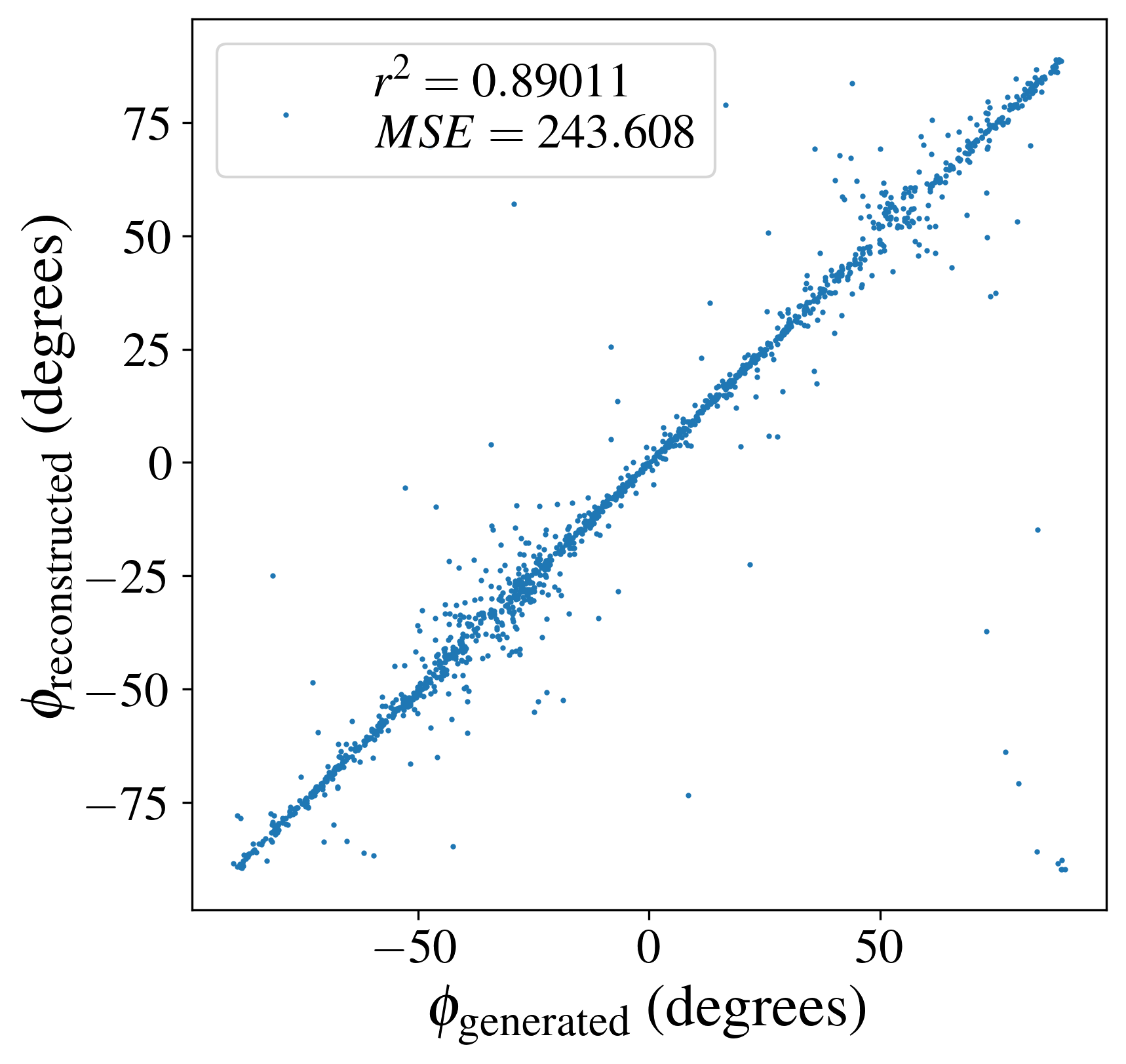}
    \caption{}
    \label{fig:azimuth}
  \end{subfigure}
  \hspace{1cm}
  \begin{subfigure}[b]{0.45\textwidth}
    \centering
    \includegraphics[height=0.8\textwidth]{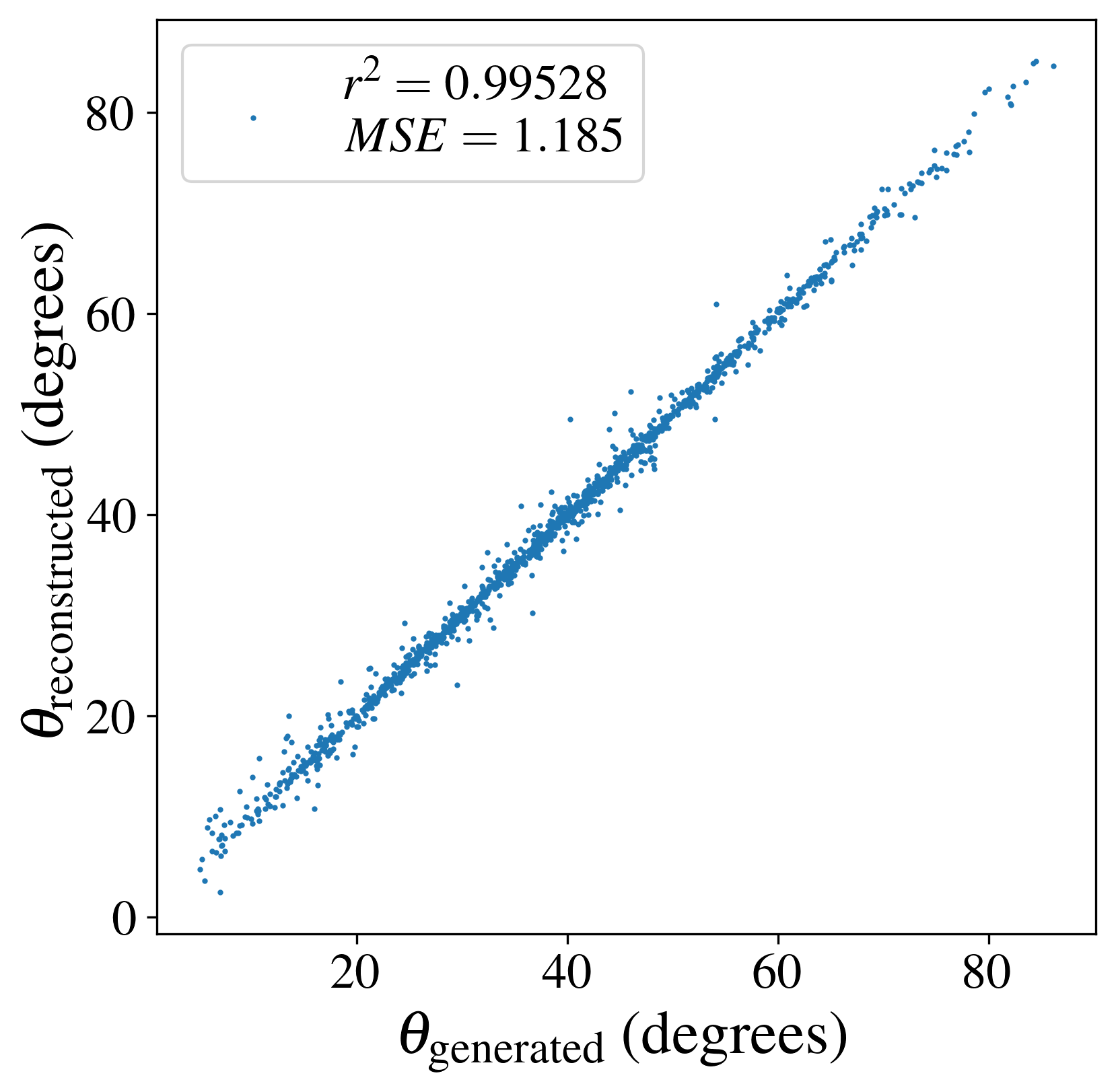}
    \caption{}
    \label{fig:zenith}
  \end{subfigure}\\
  \begin{subfigure}[b]{0.45\textwidth}
    \centering
    \includegraphics[width=.9\textwidth]{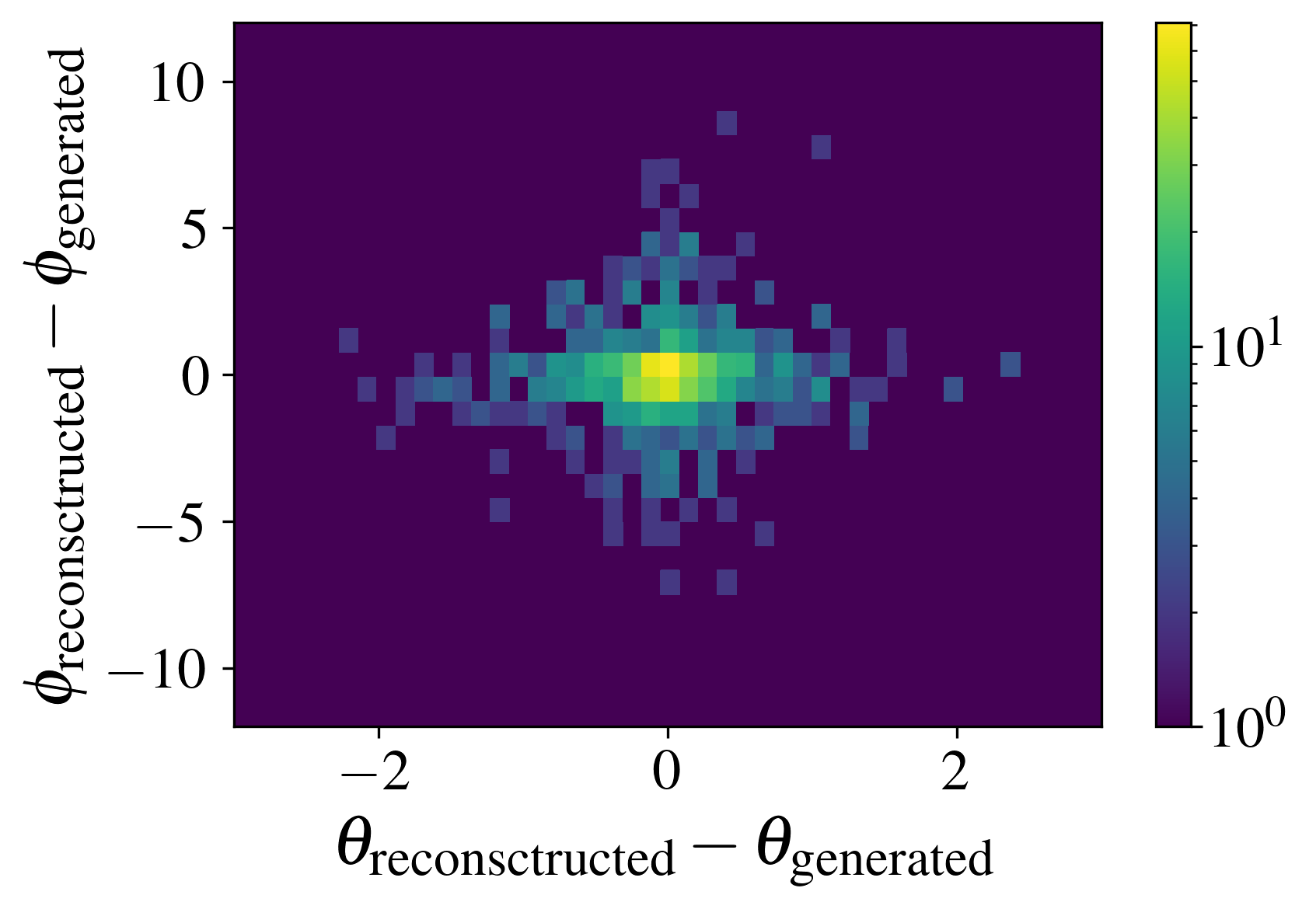}
    \caption{}
    \label{fig:zenith_azimuth}
  \end{subfigure}
  \hspace{1cm}
  \begin{subfigure}[b]{0.45\textwidth}
    \centering
    \includegraphics[width=.9\textwidth]{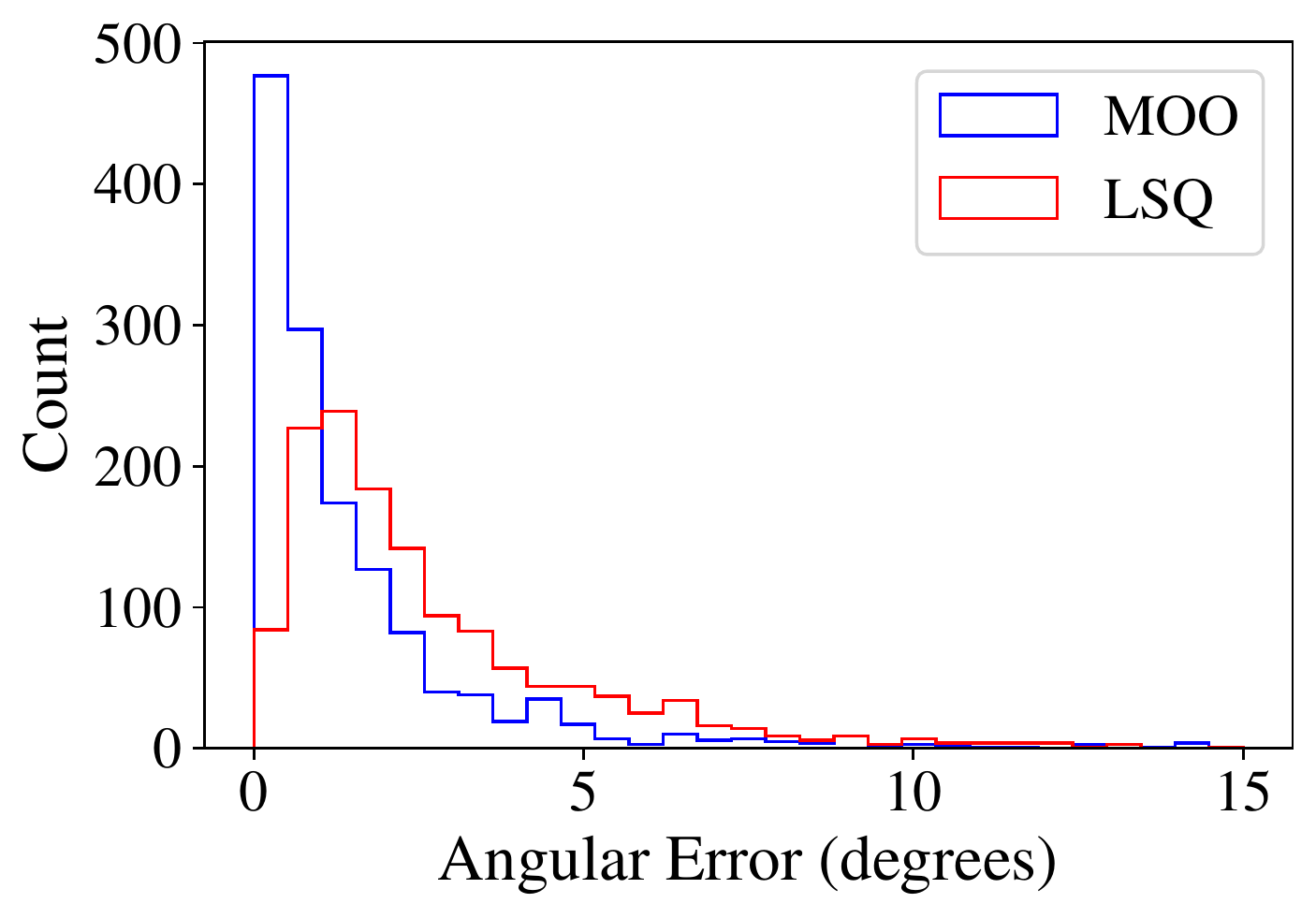}
    \caption{}
    \label{fig:pointing_accuracy}
  \end{subfigure}
  \caption{Algorithm performance of direction reconstruction. (a) Azimuthal reconstruction, (b) Zenith reconstruction, (c) Correlated azimuth vs zenith (both angles in degrees), (d) absolute pointing error. The increased scatter in the azimuthal reconstruction around $\phi\approx-30^\circ, 60^\circ$ are due to the alignment of tracks with crystal planes in the CUORE detector and are particular to the symmetries of that geometry.}
  \label{fig:angular_performance}
\end{figure}

Pointing accuracy of reconstructed trajectories is calculated by finding angular separations from the corresponding true trajectory. Fig.~\ref{fig:pointing_accuracy} shows a comparison of distributions of pointing accuracy between our algorithm and least squares. The average angular error was reduced between the methods from 2.7 degrees for least squares to 1.6 degrees for \ac{MOO}. Figure~\ref{fig:angular_performance}, shows a comparison of the reconstruction accuracy in zenith and azimuthal angles, indicating an improvement mean square error by a factor of 1.7 and 4.4 in azimuthal and zenith angles respectively between \ac{LSQ} and \ac{MOO}.

\begin{figure}
    \centering
    \begin{subfigure}[b]{0.45\textwidth}
        \centering
        \includegraphics[height=0.8\textwidth]{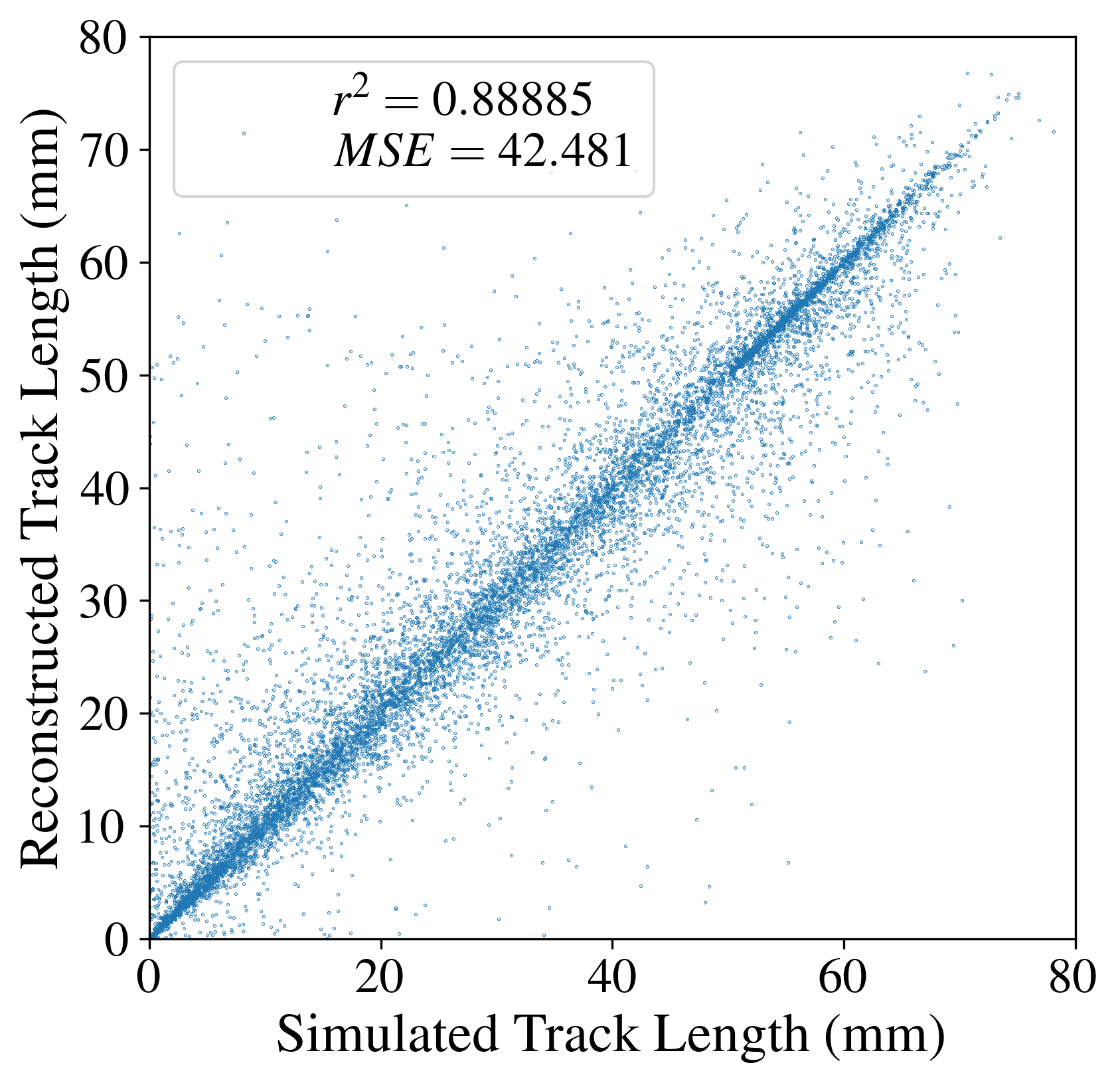}
        \caption{MOO}
        \label{fig:pathlength_moo}
    \end{subfigure}
    \hspace{1cm}
    \begin{subfigure}[b]{0.45\textwidth}
        \centering
        \includegraphics[height=0.8\textwidth]{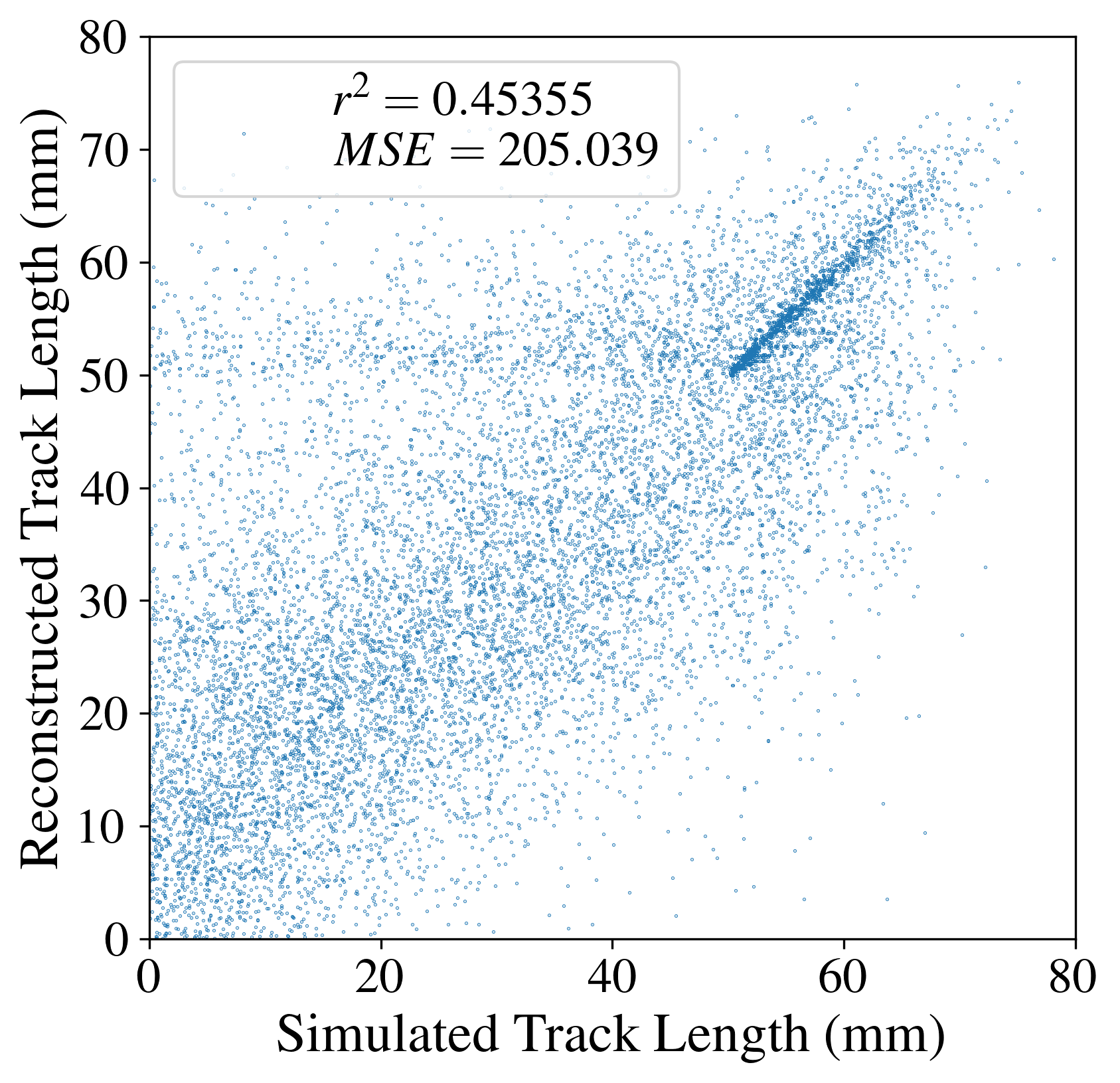}
        \caption{LSQ}
        \label{fig:pathlength_lstsq}
    \end{subfigure}
    \caption{Distribution of crystal path-length reconstructions for MOO (left) compared to LSQ (right). The tighter clustering around $y=x$ for the MOO indicates a much more reliable reconstruction of the path length through each crystal, yielding a higher correlation coefficient $r^2$ and a lower Mean Squared Error (MSE).}
    \label{fig:pathlength}
\end{figure}

Finally, we assess the ability of the algorithm to reconstruct path-lengths on the individual crystal level. The ability for this to be done accurately is of critical importance for calculating $dE/dx$ and thus for assessing the probability for a particular trajectory. This probability is taken into account in the third objective function mentioned in Sec.~\ref{sec:moo}, and could also be useful in hypothesis testing. As shown in Fig. ~\ref{fig:pathlength}, we compare the path-length reconstruction between our algorithm and \ac{LSQ} over the $\sim$12000 intersected crystals from the same sample of muons. We saw mean square error improvement by a factor of 4.8 as well as an $r^2$ increase from .45 to .89. 

\begin{figure}
    \centering
    \begin{subfigure}[b]{0.48\textwidth}
        \centering
        \includegraphics[width=0.92\textwidth]{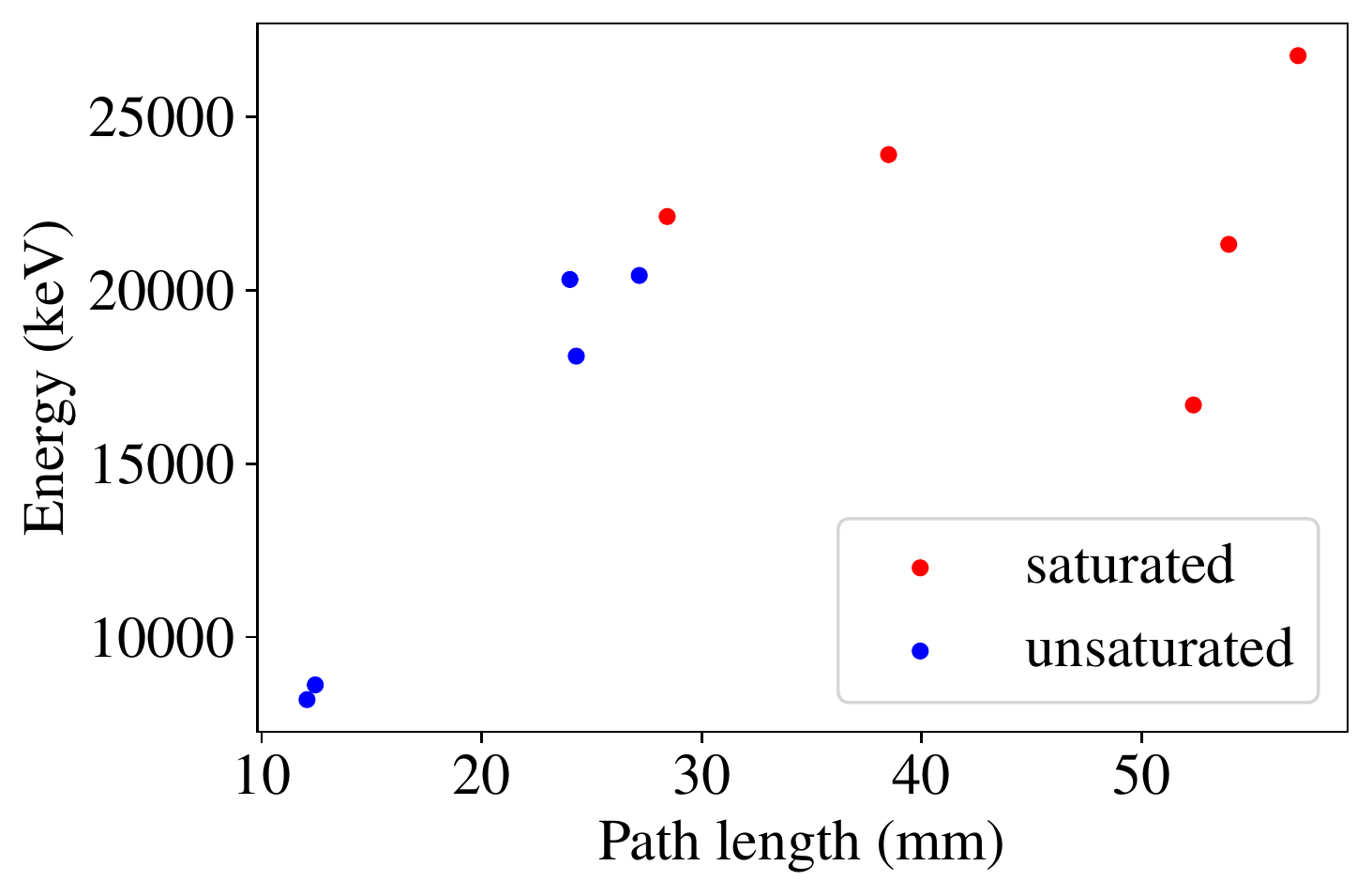}
        \caption{}
    \end{subfigure}
    \hspace{.2cm}
    \begin{subfigure}[b]{0.48\textwidth}
        \centering
        \includegraphics[width=0.92\textwidth]{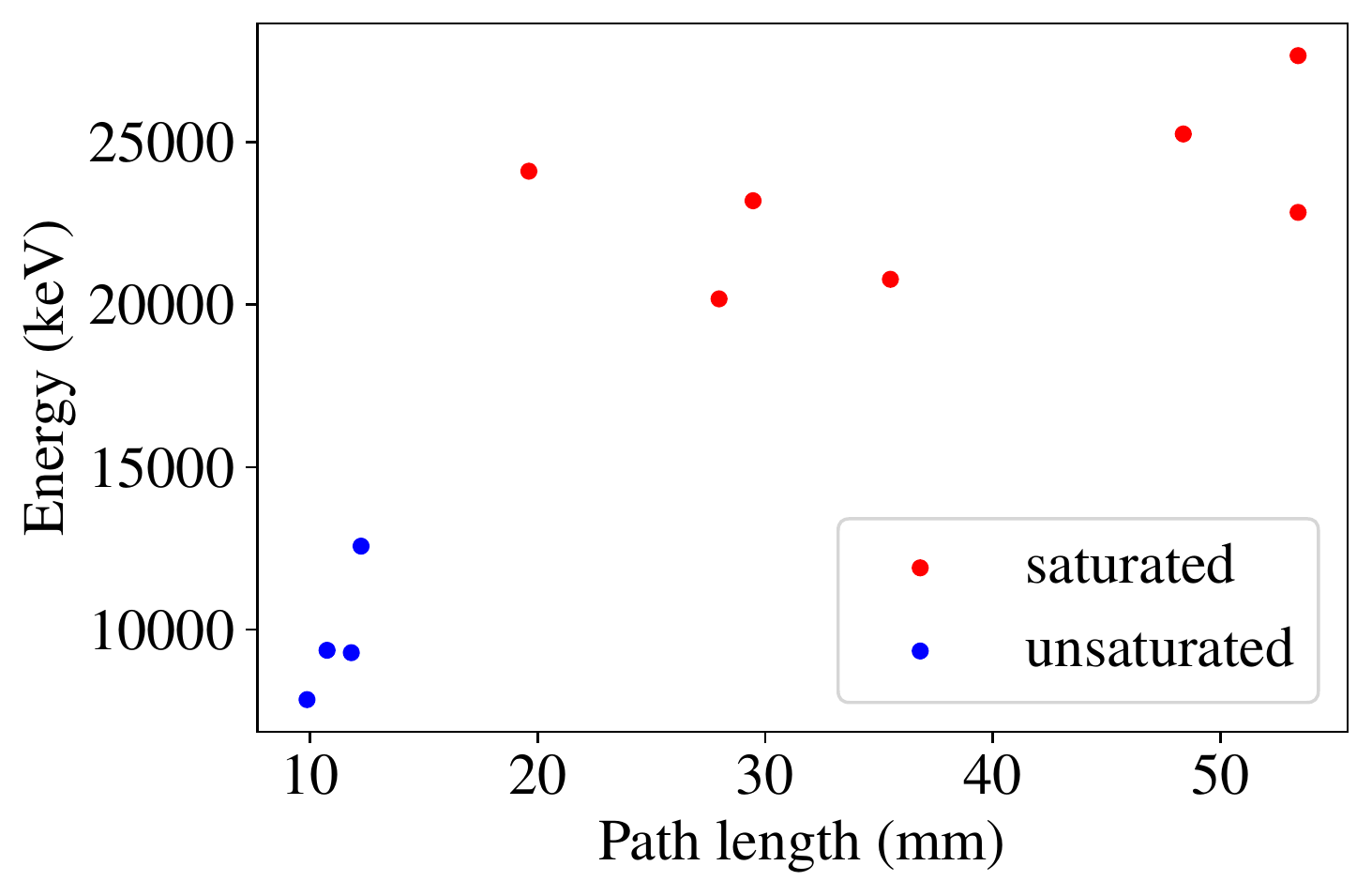}
        \caption{}
    \end{subfigure}\\
    \begin{subfigure}[b]{0.48\textwidth}
        \centering
        \includegraphics[width=0.92\textwidth]{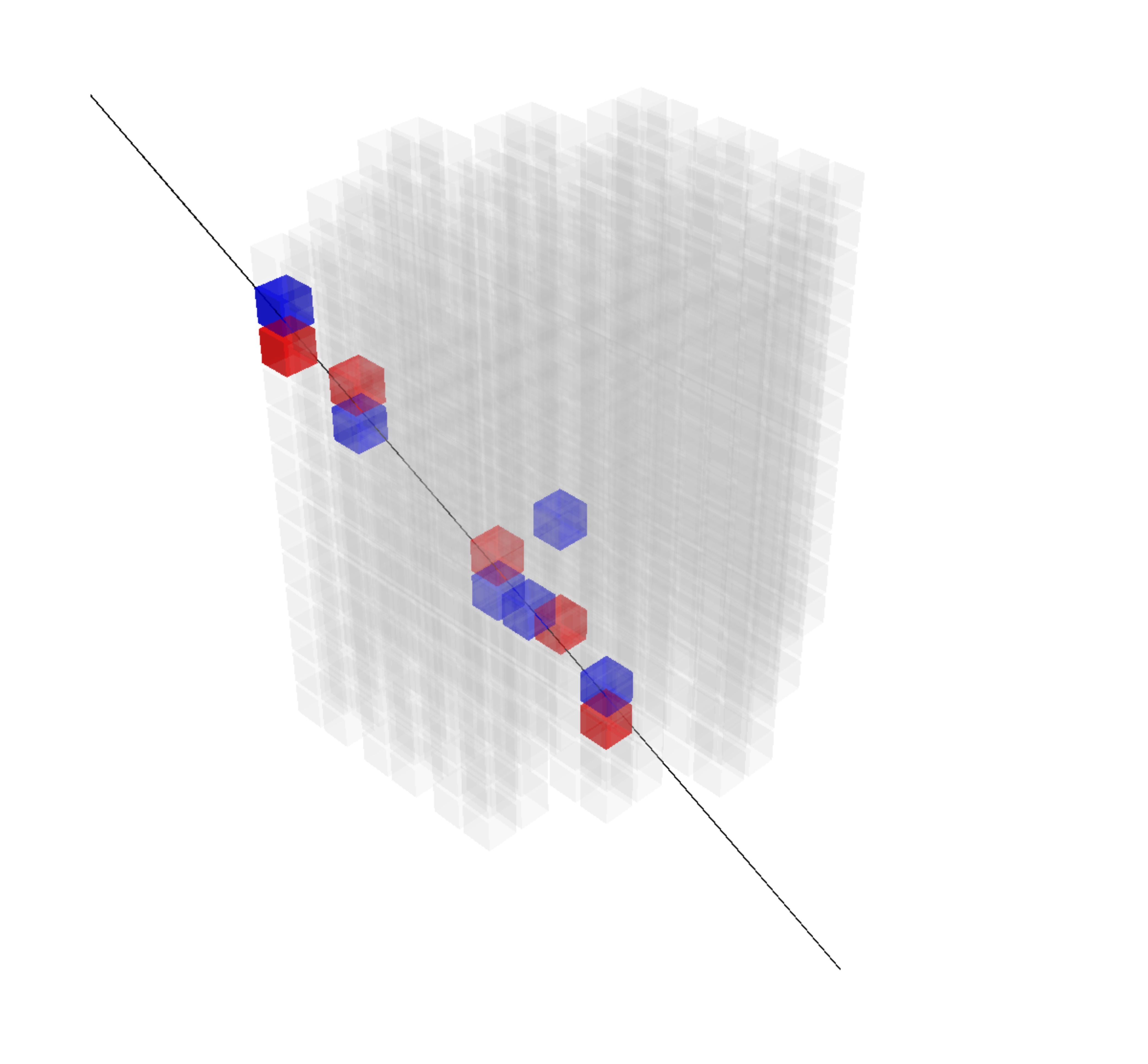}
        \caption{}
    \end{subfigure}
    \hspace{.2cm}
    \begin{subfigure}[b]{0.48\textwidth}
        \centering
        \includegraphics[width=0.92\textwidth]{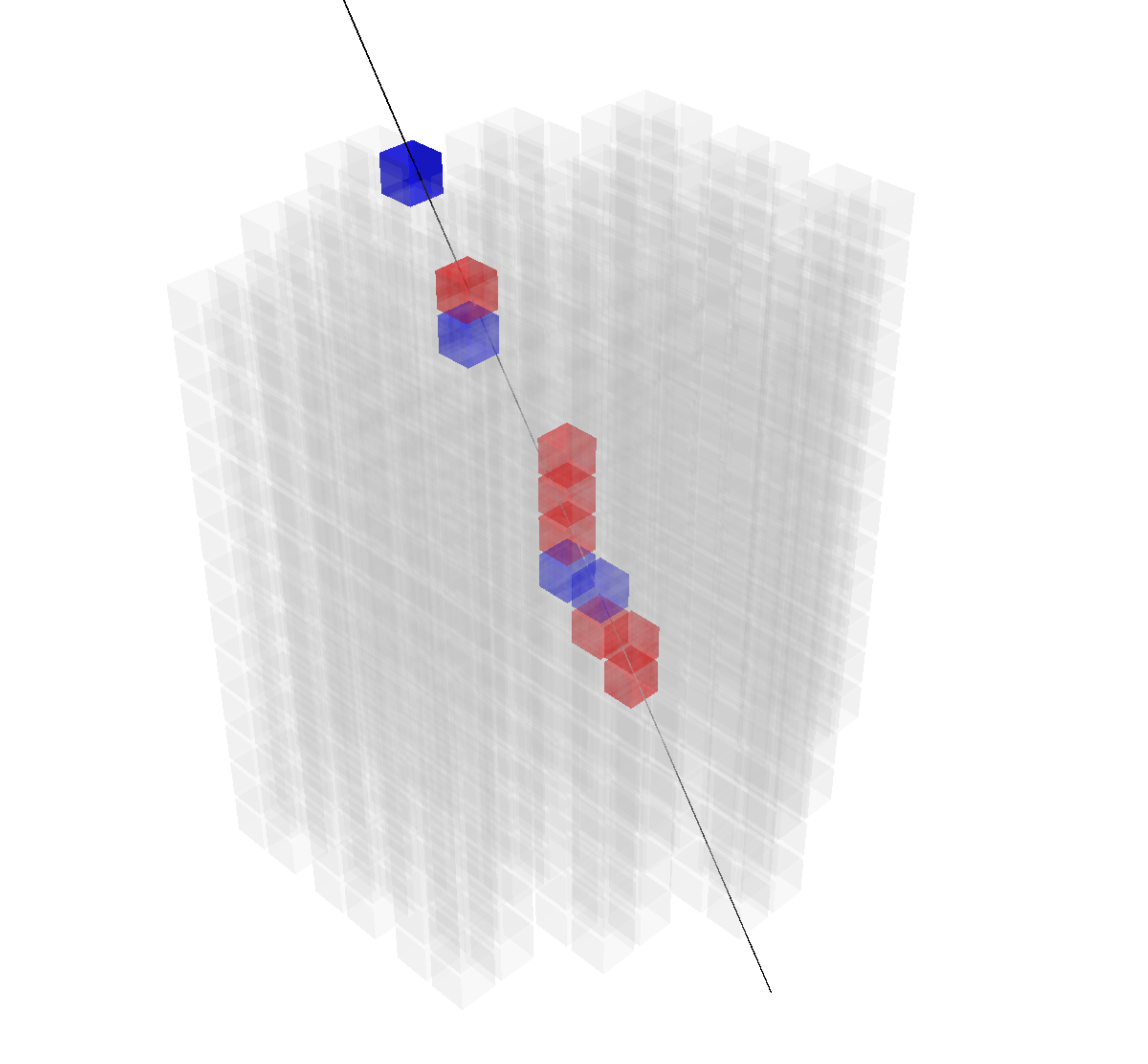}
        \caption{}
    \end{subfigure}  
    \caption{Top: Two track-like events from CUORE (left and right), that are presumed to be muons. Each point in the plot corresponds to a single lit crystal in the track. The crystal track lengths reconstructed by the MOO algorithm are plotted on the x-axis, with the reconstructed energy (from the CUORE data processing) plotted along y. The reconstructed energies are inputs into the MOO algorithm and the crystal track lengths are outputs. The saturation energy varies between the CUORE calorimeters but is typically around 20\,MeV -- above saturation the reconstructed energy can be very inaccurate. The MOO algorithm explains the calorimeter saturation by the long reconstructed track lengths. Bottom: A geometric visualization of the same tracks. The blue and red color coding corresponds to non-saturated and saturated crystals, respectively.}
    \label{fig:cuore_events}
\end{figure}

In Fig.~\ref{fig:cuore_events}, we show the algorithm performance on two track-like events from real CUORE data that are presumed to be muons. The events contain saturated pulses whose energy is poorly measured by the current reconstruction algorithms. Of course, the true track is unknown, however, the algorithm correctly explains the saturation of channels based on the reconstructed path length. This is clearly a direction for future research.

\section{Discussion}
\label{sec:discussion}

The CUORE detector is optimized for the search for neutrinoless double-beta decay, which largely treats each calorimeter individually or in an anti-coincidence mode. The algorithm presented here expands the CUORE analysis infrastructure to better reconstruct track-like muons, treating the detector holistically as a segmented detector. The crystal coarseness and lack of time-of-flight information for relativistic particles makes track-reconstruction more difficult compared to dedicated tracking detectors, nonetheless the MOO-based algorithm presented here is able to overcome such limitations in reconstructing track paths.

Comparisons to the ``naive'' \ac{LSQ}-based reconstruction highlight some of the potential issues that can manifest when fitting to coarsely segmented tracks. The \ac{LSQ} algorithm serves not only as a benchmark with which to compare the improvements of the MOO-based approach, but also indicates that the sophistication of the MOO algorithm is necessary to avoid issues arising from the discreteness of triggered channels, and eliminate the bias towards crystal centers that may generally arise when fitting to a single spatial/energy-weighted objective function.

Preliminary validation of the algorithm's performance may be evaluated within physics data collected by CUORE by looking at channel saturation within tracks. Up to fluctuations in $dE/dx$ related to the energy loss of the primary particle, energy deposition within crystals is expected to be linearly proportional to the pathlength traversed.  Channels within CUORE have typical saturation energies of 20\,MeV, meaning for that for muons with most-probably energy-depositions of $\sim10$\,MeV/cm in TeO$_2$, path-lengths through a crystal greater than $\sim2$ cm have a high probability of saturating that channel. Figure~\ref{fig:cuore_events} showing crystal energy versus reconstructed path-lengths for probably muon tracks identified within CUORE physics data, showing the expected linear relationship between path-length and energy, up until crystal saturation begins to take place for path-lengths greater than $\sim2$\,cm, as expected.

Further studies on CUORE data using this algorithm will provide a greater understanding of muons within the detector. While the muon rate within CUORE is efficiently reduced by its depth within LNGS, of particular interest is the investigation into any remaining cosmogenic activation of nuclides which can constitute backgrounds to searches for neutrinoless double beta decay.  Application of this algorithm will also be useful for the CUPID upgrade to CUORE, which plans to include a dedicated muon veto surrounding the detector \cite{CUPID-PreCDR}. 

While much of this work has considered track-like muon events, the algorithm presented can be adapted to perform searches for exotic track-like events within CUORE. It is worth noting that due to the relatively small crystal size and large fluctuations in energy deposition, it is not possible to distinguish between relativistic \ac{SM} particles such as muons, pions, and protons (see Fig.~\ref{fig:dEdx_by_particle}). But \ac{BSM} particles such as low-velocity monopoles \cite{doi:10.1098/rspa.1931.0130,PhysRev.74.817,tHooft:1974kcl,Polyakov:1974ek,PhysRevD.98.030001}, \acp{LIP} \cite{GLASHOW1961437,SCHELLEKENS1990363,CHUN1995608,Abel_2008,PhysRevD.88.015001}, and \ac{MIDM} \cite{PhysRevD.98.083516} are all predicted to produce track-like energy depositions with a substantially different $dE/dx$ compared to muons, pions, and protons.  By adjusting the $dE/dx$ \ac{PDF} within the $f_3$ objective searches for such BSM-particles can be conducted, with the likelihood ratio $f_3^{\text{BSM}}/f_3^{\text{muon}}$ being a suitable test statistic for discriminating \ac{BSM} candidates from background muons. We leave sensitivity predictions for such searches for future work.

While this algorithm was developed with the CUORE detector and geometry in mind, it is important to note that it is easily transferred to other segmented detector arrays.  The logistic regression hyperparameters $\alpha$ and $\lambda$ within the ``hit'' and ``missed'' channel objective functions may easily be optimized over using simulated data for an arbitrary detector geometry, while the rest of the algorithm is based on the full geometric implementation of the detector array.  To recall, such hyperparameters are necessary for smoothing the objective functions while the algorithm evolves to find the Pareto front.  Importantly, they are not used in the final track selection across the Pareto front, which instead utilizes the full analytic calculation of crystal-track intersections.

Other future directions for improvement to the algorithm presented here include extracting event-level uncertainties and accounting for correlated nearby channels. Event-level uncertainties could be extracted from one or more of the cost functions, but this is made slightly complicated by the fact the \ac{MOO} by definition does not combine the costs into a single function that could act as a ``likelihood''. Muons and other high energy events can produce secondaries that could deposit energy into channels that are not directly intersected by the track. This could include low-energy secondaries like Bremsstrahlung photons or high-energy ones like annihilation $\gamma$'s. These correlated off-track channels are currently treated in the same manner as an accidental coincidence, but could be incorporated into the cost functions in the future to further improve event reconstruction. Finally, the algorithm could potentially be generalized to fit multiple simultaneous tracks e.g.\ from muon shower events.

\acknowledgments

The authors would like to thank the CUORE Collaboration for useful conversations, in particular, Jorge Torres, Erin Hansen, Douglas Adams and Thomas O'Donnell. Further, we thank the CUORE Collaboration for the use of the CUORE muon events, which made use of the \texttt{DIANA} data analysis and \texttt{APOLLO} data acquisition software which has been developed by the Cuoricino, CUORE, LUCIFER, CUPID-0, and CUPID-Mo Collaborations.
This work was funded by the US Department of Energy, Office of Science, Office of Nuclear Physics under grant number DE-SC0011091.
 
\bibliographystyle{JHEP}

\providecommand{\href}[2]{#2}\begingroup\raggedright\endgroup

\end{document}